\documentclass[aps,amsmath,amssymb,showpacs,superscriptaddress,floatfix,preprint]{revtex4-1}
\usepackage{color}
\usepackage{graphicx,epsfig}
\usepackage[hypertex]{hyperref}
\DeclareMathOperator{\Tr}{Tr}
\DeclareMathOperator{\RS}{\mathrm{\scriptscriptstyle RS}}\DeclareMathOperator{\1RSB}{\mathrm{\scriptscriptstyle 1RSB}}

\begin{document}

\title{Orientational glass: full replica symmetry breaking in generalized spin glass-like models without reflection symmetry}

\author{E. E. Tareyeva}
\affiliation{Institute for High Pressure Physics, Russian Academy of Sciences, 142190, Troitsk,
 Russia}
\author{T. I. Schelkacheva}
\affiliation{Institute for High Pressure Physics, Russian Academy of Sciences, 142190, Troitsk,
 Russia}
\author{N. M. Chtchelkatchev}
\affiliation{Institute for High Pressure Physics, Russian Academy of Sciences, 142190, Troitsk,
 Russia}
\affiliation{L.D. Landau Institute for Theoretical Physics, Russian Academy of Sciences,
  117940 Moscow, Russia}
\affiliation{Department of Theoretical Physics, Moscow Institute of Physics and
Technology, 141700 Moscow, Russia}

\begin{abstract}
We investigate near  the point of glass transition the expansion of the free energy corresponding to the generalized Sherrington--Kirkpatrick model with arbitrary diagonal
operators $\hat{U}$ standing instead of Ising spins. We focus on the case when $\hat{U}$ is an operator with broken reflection symmetry. Such a consideration is important for understanding the behavior of spin-glass-like phases in a number of real physical systems, mainly in orientational glasses in mixed molecular crystals which present just the case.
We build explicitly a full replica symmetry breaking (FRSB) solution of the equations for the orientational glass order parameters when the non-symmetric part of $\hat U$ is small.
This particular result presents a counterexample in the context of usually adopted conjecture of the absence of FRSB solution in systems with no reflection symmetry.
\end{abstract}


\maketitle


\section{Introduction\label{Sec:Intro}}

The theory of spin glasses has been developed  as an attempt to describe
unordered equilibrium freezing of spins in actual dilute magnetic systems
with disorder and frustration. This problem was soon solved at the
mean-field level~\cite{EA,sk,A-T,par1,par2} [see also Ref.~\cite{book} for a review]. Below the
Almeida -- Thouless line~\cite{A-T} the replica symmetric solution was
shown to be incorrect. Parisi proposed the method of replica symmetry
breaking (RSB) step by step with the limit --- full RSB (FRSB) when glass
order parameter becomes a continuous non-decreasing function $q(x)$ of a
parameter $0\leq x\leq 1$. It provides the hierarchical distribution of
pure states overlaps probability $P(q)$ through $P(q) = dx/dq$. This
approach allows to describe the main features of the experiments on spin
glasses.

A number  of generalizations of the Sherrington and Kirkpatrick (SK) model
with Ising spins have been considered. But still the problem how  RSB
and FRSB solutions do occur remains relevant and far from being completely
understood~\cite{Oppermann,A.
Klic,Dominicis,Machta,Tem,Temesvari,T.Temesvari}. It was shown that the
violation of the replica symmetry is correlated with the symmetry
properties of the Hamiltonian~\cite{GS,nob,book1,Marin,Guerra,Leuzzi}.
There is a conjecture that in the absence of the reflection symmetry it is
not possible to construct a continuous non-decreasing function $q(x)$ and
so, the FRSB solution does not exist in this case, or at least does not
occur in the point of RS solution instability. It is possible that in some
of these models different stages of replica symmetry breaking occur not in
the first symmetry breaking point as in the SK model (see, for
example,Refs.~\cite{GKS,Gribova}, where the Potts model is
considered).

In the literature the absence of reflection (or time reversal) symmetry usually was
incorporated in the structure of the Hamiltonian. It is so, for example,
for Potts spin glasses and for $p$-spin spin glass with the interaction of
$p$ Ising spins. However, there is another way to break reflection
symmetry -- that is to incorporate the breaking in the character of the
operators themselves while the structure of the Hamiltonian remains
two--particle as in the SK model. As far as we know the present paper is
the first one to deal with RSB in such a situation.

It is worth to notice that, while, say, $p$-spin model has a physical meaning only as a prototype of structural glasses, there exists a
number of real physical systems which can be described in terms of two--operator random systems \cite{GS,PRE,Prb} of the SK type where the absence of reflection
symmetry is  caused by the characteristics of the operators $U$ themselves.

Let us list some physical examples of such systems.  They are just described by the Hamiltonians of the SK type but with Ising spins changed for non reflection symmetrical operators $\hat U$. The investigation of replica symmetry breaking presented in this paper can give some new information about low-temperature behavior of these systems.

For example $\mathrm{ortho-para}$--hydrogen mixed crystals and in $Ar-N_2$  present mixtures of spherically symmetric molecules and momentum--bearing molecules. The corresponding orientational quadrupolar  glass was investigated  on the base of the S-K type  Hamiltonian with $U=Q$, where $Q=3 J^2_{z} - 2$,  ${\bf J}=1$~\cite{Luchinskaya}. Another example of a SG--like phase in molecular crystal is  presented by pure $\mathrm{para}-H_2$ (or $\mathrm{ortho}-D_2$) under pressure. The possibility of orientational order in systems of initially spherically symmetric molecule states is due to the involving of higher order orbital moments $J = 2, 4...$ in the physics under pressure. The frustration and disorder give the basis to the investigation of quadrupole glass with $J=2$. Now $\hat U=(1/3)(3 J_z^2 -6)$. Such a theory was constructed in Ref.~\onlinecite{PRE}. Another interesting example of spin-glass-like phase is presented by the orientational glass that appears in molecular  $C_{60}$ at $90 K$. Now the sample is not a mixture and the form of the molecules is almost spherical. Nevertheless the existing anisotropy of the potential causes a frustration. The dependence of the anisotropic part of the potential on mutual orientation of molecules has two pronounced minima and so the effect of a mixture is obtained. This is the base of the theory with $\hat U$ being a superposition of cubical harmonics~\cite{C60}.

The aim of  this work -- to find out for which class of $\hat U$-operators that do not have the reflection symmetry the problem can be described by
the Parisi FRSB scheme. In this paper, we do the first step in this direction. In Ref.~\cite{Full} it was shown that in the case of
arbitrary $\hat{U}$-operators satisfying the condition of the reflection symmetry there is full replica symmetry breaking. Keeping in mind that the
FRSB solution is valid for operators with the reflection symmetry we seek for the FRSB solution for operators $\hat{U}$ represented by the sum of a
symmetric operator and a small perturbation that has no reflection symmetry. This investigation is close in spirit to the problem of building
the FRSB solution for the SK model in a weak field~\cite{par2}.

\section{The model}
We start with the Hamiltonian
\begin{equation}\label{SK}
H=-\frac{1}{2}\sum_{i\neq j}J_{ij} \hat{U}_i \hat{U}_j.
\end{equation}
where arbitrary diagonal operators $\hat{U}$ are located on  the lattice
sites $i$. The quenched interactions $J_{ij}$ are distributed with the
Gaussian probability
\begin{equation}
P(J_{i,j})=\frac{\sqrt{N}}{\sqrt{2\pi}
J}\exp\left[-\frac{(J_{i,j})^{2}N}{ 2J^{2}}\right], \label{two}
\end{equation}
where $N$ is the number of sites.
Using replica approach we can write in standard way the free energy
averaged over disorder in the form
\begin{gather}\label{free}
\langle F\rangle_J/NkT=\lim_{n \rightarrow 0}\frac{1}{n}\max\left \{ \frac{t^2}{4}\sum_{\alpha}
(w^{\alpha})^{2} + \frac{t^2}{2}\sum_{\alpha>\beta} (q^{\alpha\beta})^{2}-
\ln\Tr_{\{U^{\alpha}\}}\exp \hat{\theta}\right\}.
\end{gather}
Here
\begin{equation}
\hat{\theta}=t^2
\sum_{\alpha>\beta}(q^{\alpha\beta})\hat{U}^{\alpha}\hat{U}^{\beta}+\frac{t^2}{2}
 \sum_{\alpha}{(w^{\alpha})}(\hat{U}^{\alpha})^2, \label{six}
 \end{equation}
where $t={J}/kT$. The saddle point conditions were used
to define the glass order parameter
\begin{equation}\label{qdef}
q^{\alpha\beta}={\Tr\left[\hat{U}^{\alpha}\hat{U}^{\beta} \exp\left(\hat{\theta}\right)\right]}/{\Tr\left[\exp\left(\hat{\theta}\right)\right]},
\end{equation}
and the auxiliary order parameter
\begin{equation}\label{wdef}
 w^{\alpha}={\Tr\left[(\hat{U}^{\alpha})^2 \exp\left(\hat{\theta}\right)\right]}/{\Tr\left[\exp\left(\hat{\theta}\right)\right]}.
\end{equation}

In the RS approximation from the extremum condition for the free energy
for the glass order parameter $q_{\RS}$ we have:
\begin{equation}\label{0qrs}
q_{\RS}=\int dz^G\left\{ \frac{\Tr\left[\hat{U} \exp\left(\hat{\theta}_{\RS}\right)\right]}
{\Tr\left[\exp\left(\hat{\theta}_{\RS}\right)\right]}\right\}^{2}.
\end{equation}
and for the auxiliary order parameter:
\begin{equation}\label{00qrs}
w_{\rm {\scriptscriptstyle RS}}=\int dz^G \frac{\Tr\left[{\hat{U}}^2 \exp\left(\hat{\theta}_{\RS}\right)\right]}
{\Tr\left[\exp\left(\hat{\theta}_{\RS}\right)\right]}.
\end{equation}
Here
\begin{equation}\label{1qrs}
\hat{\theta}_{\RS}=zt\sqrt{{q_{\RS}}}\,\hat{U}+t^2\frac{[{w_{\RS}}-
{q_{\RS}}]}{2}\hat{U}^2
\end{equation}
and
\begin{equation}
\int dz^G = \int_{-\infty}^{\infty} \frac{dz}{\sqrt{2\pi}}\exp\left(-\frac{z^2}{2}\right).
\end{equation}

The stability of the RS solution can be tested by  the
investigation of the gaussian fluctuations contribution to the free energy
near this solution. The solution is stable while
all the eigen modes of the fluctuation propagator are positive. The most
important mode is the so-called replicone mode \cite{A-T,ESc} since only
its sign is usually sensitive to the replica symmetry breaking degree
and to the temperature. For example, the replica symmetric solution is
stable unless the corresponding replicon mode energy ${\lambda_{\rm(\RS)\,
repl}}>0$. The RS-solution can break at the temperature $T_{c}$
determined by the equation $\lambda_{\rm(\RS)\, repl}=0$, where
\begin{gather}\label{lambdaRS}
\lambda_{\rm (\RS)\, repl}=1 - t^{2}
\int dz^G \left\{\frac{\Tr\left(\hat{U}^2
e^{\hat{\theta}_{\RS}}\right)} {\Tr e^{\hat{\theta}_{\RS}}}-
\left[\frac{\Tr\hat{U} e^{\hat{\theta}_{\RS}}}
{\Tr e^{\hat{\theta}_{\RS}}}\right]^2\right\}^2.
\end{gather}
The equation $\lambda_{\rm(\RS)\, repl}=0$ is nothing else that the
bifurcation condition for 1RSB equation for glass order
parameter~\cite{Prb,schelk}, i.e., as the condition that a small solution
with 1RSB can appear. Analogously the other $\lambda_{(n-1)RSB}$ and the
bifurcation condition at the $n$-th stages of the replica symmetry
breaking are related.

We break the RS once again and obtain the corresponding
expressions for the free energy and the order parameters. The bifurcation
condition $\lambda_{\rm(\1RSB)\, repl}=0$ determining the
temperature follows from the condition that a nontrivial
small solution for the 2RSB glass order parameter
appears.~\cite{Prb,ESc,schelk} We have:
\begin{gather}
\lambda_{\rm(\1RSB)\, repl}= 1 - t^{2}
\int dz^G \frac{\int ds^G\left[\Tr\exp\hat{\theta}_{\1RSB}\right]^{m}
\left\{\frac{\Tr\left[\hat{U}^2
\exp\hat{\theta}_{\1RSB}\right]} {\Tr\left[\exp\hat{\theta}_{\1RSB}\right]}-
\left[\frac{\Tr\left[\hat{U} \exp\hat{\theta}_{\1RSB}\right]}
{\Tr\left[\exp\hat{\theta}_{\1RSB}\right]}\right]^2\right\}^2}{\int ds^G
\left[\Tr\exp\hat{\theta}_{\1RSB}\right]^{m}}\,,\label{lambda}
\end{gather}
where $\theta_{\1RSB}$ is the analog of (\ref{1qrs}) for the
first stage of RSB and $m$ - 1RSB order parameter.

\section{FRSB solution}

Let us consider first a generalized model  defined
by the Hamiltonian (\ref{SK}) with reflection symmetrical operators $U$.
The reflection symmetry implies that for any integer $k$,
\begin{equation}
\Tr\left[{\hat{U}}^{(2k+1)}\right]=0. \label{odd}
\end{equation}
In the RS-approximation we find the solution $q_{\RS}$ that is zero  at
high temperature. The bifurcation condition in this case is:
\begin{equation} \label{ten}
1-t_c^2 w_{\RS}^2(t_c) = 0.
\end{equation}
This equation coincides with $\lambda_{\rm(\RS)\, repl} = 0$ [see,
 e.g., Ref.~\onlinecite{book}]. It is zero high temperature solution that
bifurcates. At $T<T_c$ certain nontrivial 1RSB solutions appear but they
are unstable.

Investigating  $1$RSB, $2$RSB, $3$RSB, ..., $n$RSB,  and so on,  we see
that the equations for the glass order parameters always contain the
quantity
\begin{gather}
{\Tr[U \exp (\theta_{n{\rm\scriptscriptstyle RSB}})]}/{\Tr[\exp (\theta_{n{\rm\scriptscriptstyle RSB}})]}.
\end{gather}
Here  $\theta_{n{\rm\scriptscriptstyle RSB}}$ are the analogs of
(\ref{1qrs}) for higher stages of RSB (see Ref.~\onlinecite{Full} for
details). Therefore, one of the solutions of this equation is trivial at
each RSB-step, and the appearance of the $n$RSB solution can be regarded
as the bifurcation of the trivial $(n-1)$RSB solution. In this case, the
equation  $\lambda _{n{\rm\scriptscriptstyle RSB}} = 0$ coincides with the
corresponding branching condition (\ref{ten}). This means that in any
case, the $n$RSB solutions at different stages of the symmetry breaking
can exist at temperature $T< T_{c}$ determined by this bifurcation
condition, and so we always can look for $\mathrm{FRSB}$ solution. Writing
the free energy as a series over $\delta q^{\alpha\beta }$ near $T_c$ (up
to the fourth order)  we obtain $q(x)=cx$ in the leading approximation [a
similar procedure was described in details in Ref.~\onlinecite{Full}].

If the operators $\hat{U}$ do not have the reflection symmetry, $
\Tr{\hat{U}}^{(2k+1)} \neq 0 $, then the glass freezing scenario is
different from the previous case. The characteristic properties of system
develop themselves already in the replica symmetry approximation. The
nonlinear integral equation for the RS-glass order parameter simply has no
trivial solutions at any temperature because the integrand is nonsymmetric
due to the cubic terms in the free-energy expansion~\cite{Prb,ESc}. There
is a smooth increase in the RS order parameters  as the temperature
decreases. The bifurcation condition $\lambda_{\rm(\RS)\, repl}=0$
(\ref{lambdaRS}) defines the point $T_{c}$ where the RS-solution becomes
unstable.

Similarly to the previous case, considering respectively 1RSB,
2RSB, 3RSB, ...,
$n\mathrm{RSB}$, we find that the equation
$\lambda_{n{\rm\scriptscriptstyle RSB}} = 0$  always has the solution
which determines the point $T_{c}$ and coincides with the solution of
equation $\lambda_{\rm(\RS)\, repl}=0$~\cite{Prb,ESc,schelk}.  It is important
that it is the non-zero solution that bifurcates.

Looking now in general at the free--energy series over  the glass order
parameter we see that the series contain explicitly the terms which can be
classified by the reflection symmetry.

To estimate the form of the FRSB-solution near the  bifurcation  point
$T_c$ at which it ceases to coincide with the RS-solution (i.e., in
neighborhood of $T_c$), we expand the expression for the free energy up to
the fourth  order (inclusively),  assuming that the deviations $\delta
q^{\alpha \beta}$ from $q_{\RS}$  are small. We believe that one can
neglect the changes of the order parameter $w_{\RS}$ (see
Ref.~\onlinecite{schelk} where it is directly shown for 1RSB). We obtain
the deviation $\Delta F$ of the free energy  from its RS
part:
\begin{multline}\label{10frs}
\frac{\Delta F}{NkT}=\lim_{n \rightarrow 0}\frac{1}{n}\Biggl\{\frac{t^2}{4}\left[1-t^{2}W\right]
{\sum_{\alpha,\beta}}^{'}\left(\delta q^{\alpha\beta}\right)^{2}-\frac{t^{4}}{2}L{\sum_{\alpha,\beta,\delta}}^{'}\delta q^{\alpha\beta}\delta
q^{\alpha\delta}- t^{6}\biggl[B_{2}{\sum_{\alpha,\beta,\gamma,\delta}}^{'}\delta q^{\alpha\beta}\delta q^{\alpha\gamma}\delta q^{\beta\delta}+
\\
B'_{2}{\sum_{\alpha,\beta,\gamma,\delta}}^{'}\delta q^{\alpha\beta}\delta q^{\alpha\gamma}\delta q^{\alpha\delta}+
B_{3}{\sum_{\alpha,\beta,\gamma}}^{'}\delta q^{\alpha\beta}\delta q^{\beta\gamma}\delta q^{\gamma\alpha}+B'_{3}{\sum_{\alpha,\beta,\gamma}}^{'}\left(\delta
q^{\alpha\beta}\right)^{2}\delta q^{\alpha\gamma}+B_{4}{\sum_{\alpha,\beta}}^{'}\left(\delta q^{\alpha\beta}\right)^{3}\biggr]+
\\
t^{8}\biggl[D_{2}{\sum_{\alpha,\beta}}^{'}\left(\delta q^{\alpha\beta}\right)^{4}+D_{31}{\sum_{\alpha,\beta,\gamma}}^{'}\left(\delta q^{\alpha\beta}\right)^{3}\delta q^{\alpha\gamma} + D_{32}{\sum_{\alpha,\beta,\delta}}^{'}(\delta q^{\alpha\beta})^{2}\left(\delta q^{\alpha\delta}\right)^{2}+D_{33}{\sum_{\alpha,\beta,\gamma}}^{'}\left(\delta
q^{\alpha\beta}\right)^{2}\delta q^{\alpha\gamma}\delta q^{\gamma\beta}+
\\
D_{42}{\sum_{\alpha,\beta,\gamma,\delta}}^{'}\left(\delta q^{\alpha\beta}\right)^{2}\delta q^{\alpha\gamma}\delta q^{\alpha\delta}+D_{43}{\sum_{\alpha,\beta,\gamma,\delta}}^{'}
\left(\delta q^{\alpha\beta}\right)^{2}\delta q^{\alpha\gamma}\delta q^{\beta\delta}+
D_{45}{\sum_{\alpha,\beta,\gamma,\delta}}^{'}\left(\delta q^{\alpha\beta}\right)^{2}\delta q^{\alpha\gamma}\delta q^{\gamma\delta}+
\\
D_{46}{\sum_{\alpha,\beta,\gamma,\delta}}^{'}\delta q^{\alpha\beta}\delta q^{\alpha\gamma}\delta q^{\alpha\delta}\delta q^{\beta\gamma}+
D_{47}{\sum_{\alpha,\beta,\gamma,\delta}}^{'}\delta q^{\alpha\beta}\delta q^{\beta\gamma}\delta q^{\gamma\delta}\delta q^{\delta\alpha}+D_{53}{\sum_{\alpha,\beta,\gamma,\delta,\mu}}^{'}\delta q^{\alpha\beta}\delta q^{\alpha\gamma}\delta q^{\alpha\delta}q^{\alpha\mu}+
\\
D_{54}{\sum_{\alpha,\beta,\gamma,\delta,\mu}}^{'}\delta q^{\alpha\beta}\delta q^{\alpha\gamma}\delta q^{\alpha\delta}q^{\beta\mu}+
D_{55}{\sum_{\alpha,\beta,\gamma,\delta,\mu}}^{'}\delta
q^{\alpha\beta}\delta q^{\alpha\gamma}\delta q^{\gamma\delta}q^{\delta\mu}\biggr]\Biggr\},
\end{multline}
where $t=t_{c}+\Delta t$. The prime on the sum  means that only the
superscripts belonging to the same  $\delta q$ are necessarily different
in  $\sum'$.  The expressions for coefficients $W, L,...,D_{}$ are given
in Appendix. All coefficients depend only on RS-solution at $T_{c}$. Note
that $1-t^{2}W = \lambda_{\rm(\RS)\, repl}$. The expression (\ref{10frs})
includes a part without the reflection symmetry, namely, the terms with
odd number of identical replica indices (see also ~\cite{Tem, Temesvari,T.Temesvari}).

To write the free energy as a  functional of $q (x)$ we use the standard formalized algebra
rules~\cite{par2, book}. The properties of this algebra were formulated by Parisi for Ising
spin glasses. In our case, the expansion of the generalized expression
for the free energy Eq.(\ref{10frs}) includes some terms of non-standard
form. Those terms are not formally described by the Parisi rules, but can
be easily reduced to the standard form.To do this, we compared the corresponding expression, consistently producing 1RSB, 2RSB, ... symmetry breaking.
For example,
\begin{gather}\label{10001eq:JP}
\lim_{n \rightarrow
0}\frac{1}{n}{\sum_{\alpha,\beta,\gamma,\delta}}^{'}\delta
q^{\alpha\beta}\delta q^{\alpha\gamma}\delta q^{\alpha\delta}=
\lim_{n \rightarrow
0}\frac{1}{n}{\sum_{\alpha,\beta,\gamma,\delta}}^{'}\delta
q^{\alpha\beta}\delta q^{\alpha\gamma}\delta q^{\beta\delta}
\end{gather}

The equation for order parameter follows from the  stationarity condition
$(\delta/\delta q(x))\Delta F = 0$ applied to the free energy functional.
The resulting complicated integral stationarity equation can be simplified
using the differential operator $\hat{O}=
\frac{1}{q'}\frac{d}{dx}\frac{1}{q'}\frac{d}{dx}$, where $q'= \frac{d
q(x)}{dx}$. As a result, we obtain:
\begin{multline}\label{ten7888}
t^{6}\left\{B_{4}-B_{3}x\right\}+t^{8}\Biggl\{-D_{46}x\langle q \rangle+
\\
D_{47}\biggl[-4q(x)x^{2}-4x\langle q \rangle+4x\int_{0}^{x}dy q(y) \biggr]+D_{31}\langle q \rangle +
\\
D_{33}\biggl[4q(x)x+2\langle q \rangle-2\int_{0}^{x}dy q(y) \biggr]-4D_{2}q(x)\Biggr\}=0,
\end{multline}
where $\langle q \rangle=\int_{0}^{1}dy q(y)$. Our results agree  with
those obtained in Ref.~\onlinecite{C. De Dominicis} in the case when $ \hat
{U} $ are Ising spins and $(\hat{U})^{2}=1$.

Differentiating the equation Eq.\eqref{ten7888} we obtain in the
leading approximation:
\begin{equation}\label{ten73}
q'=\frac{B_{3}}{t^{2}4\left[-D_{2}+D_{33}x-D_{47}x^{2}\right]}.
\end{equation}

 In deriving the equation Eq.\eqref{ten73} we have neglected in the numerator of the members terms of the form $ q(x) \sim \tau = (T_{c}-T)/T_{c}$ compared with the constant $B_{3}$.  This is true for the consideration of Parisi~\cite{par1,par2,book}.  Since the model with the operators $\hat{U}$ with reflective symmetry ( i.e.  $\Tr\left[{\hat{U}}^{(2k+1)}\right]=0$) exactly the same as the model of the Ising spins~\cite{Full}, for reasons of continuity, we consider  operators who have a very small
 $\Tr\left[{\hat{U}}^{(2k+1)}\right]$.

It is easy to see that
\begin{gather}\label{B3RS}
B_{3}=
\frac{1}{6}\int dz^G \left\{\frac{\Tr\left(\hat{U}^2
e^{\hat{\theta}_{\RS}}\right)} {\Tr e^{\hat{\theta}_{\RS}}}-
\left[\frac{\Tr\hat{U} e^{\hat{\theta}_{\RS}}}
{\Tr e^{\hat{\theta}_{\RS}}}\right]^2\right\}^3\geq0.
\end{gather}

It follows from the Cauchy--Schwarz inequality that the expression $\left(\Tr\hat{U}^{2}e^{\hat{\theta}_{\RS}}\right)\left(\Tr e^{\hat{\theta}_{\RS}}\right)\geq\left(\Tr \hat{U}e^{\hat{\theta}_{\RS}}\right)^{2}$ follows from $\left(\sum_{n}{A_{n}}^{2}\right)\left(\sum_{n}{B_{n}}^{2}\right)\geq\left(\sum_{n}A_{n}B_{n}\right)^{2} $.

We have also
\begin{multline}
-D_{2}=\frac{1}{48}
\int dz^G \left\{\frac{\Tr\left(\hat{U}^4
e^{\hat{\theta}_{\RS}}\right)} {\Tr e^{\hat{\theta}_{\RS}}}
-3\left[\frac{\Tr\hat{U}^2 e^{\hat{\theta}_{\RS}}}
{\Tr e^{\hat{\theta}_{\RS}}}\right]^2- \right.
\\
\left.
6\left[\frac{\Tr\hat{U} e^{\hat{\theta}_{\RS}}}
{\Tr e^{\hat{\theta}_{\RS}}}\right]^4+
12\left[\frac{\Tr\hat{U} e^{\hat{\theta}_{\RS}}}
{\Tr e^{\hat{\theta}_{\RS}}}\right]^2
\left[\frac{\Tr\hat{U}^2 e^{\hat{\theta}_{\RS}}}
{\Tr e^{\hat{\theta}_{\RS}}}\right]-\right.
\\
\left.
4\left[\frac{\Tr\hat{U} e^{\hat{\theta}_{\RS}}}
{\Tr e^{\hat{\theta}_{\RS}}}\right]
\left[\frac{\Tr\hat{U}^3 e^{\hat{\theta}_{\RS}}}
{\Tr e^{\hat{\theta}_{\RS}}}\right]
\right\}^2\geq0.
\end{multline}

\begin{gather}\label{D47RS}
-D_{47}=
\frac{1}{8}\int dz^G \left\{\frac{\Tr\left(\hat{U}^2
e^{\hat{\theta}_{\RS}}\right)} {\Tr e^{\hat{\theta}_{\RS}}}-
\left[\frac{\Tr\hat{U} e^{\hat{\theta}_{\RS}}}
{\Tr e^{\hat{\theta}_{\RS}}}\right]^2\right\}^4\geq0.
\end{gather}

If $ \hat {U}$ are operators with reflection symmetry, then $D_{33}=0$,
and the denominator of (\ref{ten7888}) is positive. So, in the case of
reflection symmetry $q'>0$.

If the operators $\hat {U}$ have no reflection symmetry,  then
$D_{33}\neq0$ and the denominator of (\ref{ten7888}) can be negative.
However, for reasons of continuity, one can imagine that it is not always
the case. As an example, let us consider the operators $\hat U = \hat S +
\eta \hat Q$ where $\eta$ is small. Here $\hat S$ is the $z$-component of
spin  (for ${\bf S}=1$) taking values $(0,1,-1)$. While $\hat Q$ is the
axial quadrupole moment, $\hat Q = 3 {\hat S}^2 -2$, and it takes values
$(-2, 1, 1)$ (see, e.g. \cite{Full}). The operators $\hat Q$ and $\hat S$
have the following properties: ${\hat Q}^2 = 2 - \hat Q$, $3 {\hat S}^2= 2
+ \hat Q$, and $\hat Q \hat S = \hat S \hat Q = \hat S$. So the algebra of
these operators is closed. The operator $\hat S$ has the reflection
symmetry while $\hat Q$ has not. FRSB is valid for arbitrary reflection
symmetric operators~\cite{Full}, in particular, for $\hat S$. Let us note
that the operator $\sqrt{3}S = V$ is a second component of the quadrupole
momentum operator $V=S_{x}^2 - S_{y}^2$ considered in the problem of
anisotropic quadrupolar glass.

The detailed calculation was performed and it was shown that for small $\eta$ there is the vicinity of Tc where $q'(x)$ remains to be positive. The area where $q(x)$ depends on $x$ is small
and the function $q(x)$ is small $\sim \tau = (T_{c}-T)/T_{c}$.

 Thus we can construct correct FRSB
solution with $q'>0$ in the absence of reflection symmetry of operators $\hat U $.

It is a possibility that at lower temperature the $D_{33}$ term occurs
to be larger so that the derivative (\ref{ten7888}) becomes
negative and the FRSB ceases to exist. This fact can give rise to a
``reverse'' behavior of the system as compared with
``standard'' case proposed in Ref.~\onlinecite{GKS} for Potts glass models.

\section{Conclusions \label{Sec:Conc}}
So, we have considered a model  with two-particle interaction  where the
absence of reflection symmetry is caused by the characteristics of the
operators $U$ themselves. FRSB is first described in such a system. An
expansion for the free energy of our generalized SK model with arbitrary
operators $\hat {U}$ standing instead of Ising spins  is investigated near
the RSB transition point $T_c$. The principal prescription for obtaining a
full replica symmetry breaking solution is derived in general case. In a
case when $\hat {U}$ is a reflection symmetric operator with a
nonsymmetric perturbative part the FRSB solution is constructed
explicitly.

\section{Acknowledgments}

The work was supported by Russian Foundation for Basic Research (grants  12-03-00757-a, 10-02-00882-a, 10-02-00694a, 10-02-00700 and 11-02-00-341a), Ural Division of Russian Academy of Sciences (grant RCP-12-P3) and Presidium of Russian Academy of Sciences (program № 12-P-3-1013).

\section{Appendix}

\begin{widetext}
\begin{align}\label{AARS} W=&\langle\hat{U}_{1}^2\hat{U}_{2}^2\rangle-2\langle\hat{U}_{1}^2\hat{U}_{2}\hat{U}_{3}\rangle+
 \langle\hat{U}_{1}\hat{U}_{2}\hat{U}_{3}\hat{U}_{4}\rangle=\int dz^G \left\{\frac{\Tr\left(\hat{U}^2
e^{\hat{\theta}_{\RS}}\right)} {\Tr e^{\hat{\theta}_{\RS}}}-
\left[\frac{\Tr\hat{U} e^{\hat{\theta}_{\RS}}}
{\Tr e^{\hat{\theta}_{\RS}}}\right]^2\right\}^2.
 \end{align}

Notation used below are obvious from the equation (\ref{AARS}). The coefficients of the terms of the third and the fourth order we write out only those that are included in the final equation.
 \begin{align}
 L=&\langle\hat{U}_{1}^2\hat{U}_{2}\hat{U}_{3}\rangle-
 \langle\hat{U}_{1}\hat{U}_{2}\hat{U}_{3}\hat{U}_{4}\rangle.
\end{align}

\begin{multline}
B_{4}=\frac{1}{3}\langle\hat{U}_{1}\hat{U}_{2}\hat{U}_{3}\hat{U}_{4}\hat{U}_{5}\hat{U}_{6}\rangle-
 \langle\hat{U}_{1}^{2}\hat{U}_{2}\hat{U}_{3}\hat{U}_{4}\hat{U}_{5}\rangle+\frac{1}{3}\langle\hat{U}_{1}^{3}\hat{U}_{2}\hat{U}_{3}\hat{U}_{4}\rangle+
 \\
 \frac{3}{4}\langle\hat{U}_{1}^{2}\hat{U}_{2}^{2}\hat{U}_{3}\hat{U}_{4}\rangle-\frac{1}{2}\langle\hat{U}_{1}^{3}\hat{U}_{2}^{2}\hat{U}_{3}\rangle+
 \frac{1}{12}\langle\hat{U}_{1}^{3}\hat{U}_{2}^{3}\rangle;
\end{multline}

\begin{multline}
D_{31}=-\frac{1}{6}\langle\hat{U}_{1}^{4}\hat{U}_{2}^{3}\hat{U}_{3}\rangle+
\frac{1}{2}\langle\hat{U}_{1}^{4}\hat{U}_{2}^{2}\hat{U}_{3}\hat{U}_{4}\rangle+
\frac{2}{3}\langle\hat{U}_{1}^{3}\hat{U}_{2}^{3}\hat{U}_{3}\hat{U}_{4}\rangle+
\frac{1}{2}\langle\hat{U}_{1}^{3}\hat{U}_{2}^{2}\hat{U}_{3}^{2}\hat{U}_{4}\rangle-
4\langle\hat{U}_{1}^{3}\hat{U}_{2}^{2}\hat{U}_{3}\hat{U}_{4}\hat{U}_{5}\rangle-
\\
\frac{3}{2}\langle\hat{U}_{1}^{2}\hat{U}_{2}^{2}\hat{U}_{3}^{2}\hat{U}_{4}\hat{U}_{5}\rangle-
\frac{1}{3}\langle\hat{U}_{1}^{4}\hat{U}_{2}\hat{U}_{3}\hat{U}_{4}\hat{U}_{5}\rangle+
7\langle\hat{U}_{1}^{2}\hat{U}_{2}^{2}\hat{U}_{3}\hat{U}_{4}\hat{U}_{5}\hat{U}_{6}\rangle+
\frac{7}{3}\langle\hat{U}_{1}^{3}\hat{U}_{2}\hat{U}_{3}\hat{U}_{4}\hat{U}_{5}\hat{U}_{6}\rangle-
\\
7\langle\hat{U}_{1}^{2}\hat{U}_{2}\hat{U}_{3}\hat{U}_{4}\hat{U}_{5}\hat{U}_{6}\hat{U}_{7}\rangle+
2\langle\hat{U}_{1}\hat{U}_{2}\hat{U}_{3}\hat{U}_{4}\hat{U}_{5}\hat{U}_{6}\hat{U}_{7}\hat{U}_{8}\rangle;
\end{multline}
\begin{multline}
D_{33}=-\frac{1}{4}\langle\hat{U}_{1}^{3}\hat{U}_{2}^{3}\hat{U}_{3}^{2}\rangle+
 \frac{1}{4}\langle\hat{U}_{1}^{3}\hat{U}_{2}^{3}\hat{U}_{3}\hat{U}_{4}\rangle+
\frac{3}{2}\langle\hat{U}_{1}^{3}\hat{U}_{2}^{2}\hat{U}_{3}^{2}\hat{U}_{4}\rangle-
\frac{5}{2}\langle\hat{U}_{1}^{3}\hat{U}_{2}^{2}\hat{U}_{3}\hat{U}_{4}\hat{U}_{5}\rangle-
\frac{9}{4}\langle\hat{U}_{1}^{2}\hat{U}_{2}^{2}\hat{U}_{3}^{2}\hat{U}_{4}\hat{U}_{5}\rangle+
\\
\langle\hat{U}_{1}^{3}\hat{U}_{2}\hat{U}_{3}\hat{U}_{4}\hat{U}_{5}\hat{U}_{6}\rangle+
\frac{21}{4}\langle\hat{U}_{1}^{2}\hat{U}_{2}^{2}\hat{U}_{3}\hat{U}_{4}\hat{U}_{5}\hat{U}_{6}\rangle-
4\langle\hat{U}_{1}^{2}\hat{U}_{2}\hat{U}_{3}\hat{U}_{4}\hat{U}_{5}\hat{U}_{6}\hat{U}_{7}\rangle+
\langle\hat{U}_{1}\hat{U}_{2}\hat{U}_{3}\hat{U}_{4}\hat{U}_{5}\hat{U}_{6}\hat{U}_{7}\hat{U}_{8}\rangle;
\end{multline}
\begin{multline}
D_{46}=
-\frac{1}{2}\langle\hat{U}_{1}^{3}\hat{U}_{2}^{2}\hat{U}_{3}^{2}\hat{U}_{4}\rangle+
\langle\hat{U}_{1}^{3}\hat{U}_{2}^{2}\hat{U}_{3}\hat{U}_{4}\hat{U}_{5}\rangle+
\frac{3}{2}\langle\hat{U}_{1}^{2}\hat{U}_{2}^{2}\hat{U}_{3}^{2}\hat{U}_{4}\hat{U}_{5}\rangle-
4\langle\hat{U}_{1}^{2}\hat{U}_{2}^{2}\hat{U}_{3}\hat{U}_{4}\hat{U}_{5}\hat{U}_{6}\rangle-
\\
\frac{1}{2}\langle\hat{U}_{1}^{3}\hat{U}_{2}\hat{U}_{3}\hat{U}_{4}\hat{U}_{5}\hat{U}_{6}\rangle+
\frac{7}{2}\langle\hat{U}_{1}^{2}\hat{U}_{2}\hat{U}_{3}\hat{U}_{4}\hat{U}_{5}\hat{U}_{6}\hat{U}_{7}\rangle-
\langle\hat{U}_{1}\hat{U}_{2}\hat{U}_{3}\hat{U}_{4}\hat{U}_{5}\hat{U}_{6}\hat{U}_{7}\hat{U}_{8}\rangle.
\end{multline}
\end{widetext}

\end{document}